\documentclass[conference]{IEEEtran}
\IEEEoverridecommandlockouts

\usepackage{amssymb}
\usepackage{amsmath}
\usepackage{booktabs}
\usepackage{graphicx}
\usepackage{lineno}
\usepackage{caption}
\usepackage{subcaption}
\usepackage{tabu}
\usepackage{float}
\usepackage{enumitem}
\usepackage{longtable}
\usepackage{tabularx}
\usepackage{wrapfig}
\usepackage{dblfloatfix}

\begin{document}

\title{A Practical Framework for Preventing Distracted Pedestrian-related Incidents using Wrist Wearables}

\author{
    \IEEEauthorblockN{Nisha Vinayaga-Sureshkanth\IEEEauthorrefmark{1}, Anindya Maiti\IEEEauthorrefmark{1}, Murtuza Jadliwala\IEEEauthorrefmark{1}, Kirsten Crager\IEEEauthorrefmark{2}, Jibo He\IEEEauthorrefmark{2}, Heena Rathore\IEEEauthorrefmark{3}}
    \IEEEauthorblockA{\IEEEauthorrefmark{1}University of Texas at San Antonio, Texas, USA}
    \IEEEauthorblockA{\IEEEauthorrefmark{2}Wichita State University, Kansas, USA}
    \IEEEauthorblockA{\IEEEauthorrefmark{3}Hiller Measurements, Texas, USA}
}

\maketitle
\begin{abstract}
Distracted pedestrians, like distracted drivers, are an increasingly dangerous threat and precursors to pedestrian accidents in urban communities, often resulting in grave injuries and fatalities. Mitigating such hazards to pedestrian safety requires employment of pedestrian safety systems and applications that are effective in detecting them. Designing such frameworks is possible with the availability of sophisticated mobile and wearable devices equipped with high-precision on-board sensors capable of capturing fine-grained user movements and context, especially distracted activities. However, the key technical challenge is accurate recognition of distractions with minimal resources in real-time given the computation and communication limitations of these devices. Several recently published works improve distracted pedestrian safety by leveraging on complex activity recognition frameworks using mobile and wearable sensors to detect pedestrian distractions. Their primary focus, however, was to achieve high detection accuracy, and therefore most designs are either resource intensive and unsuitable for implementation on mainstream mobile devices, or computationally slow and not useful for real-time pedestrian safety applications, or require specialized hardware and less likely to be adopted by most users. In the quest for a pedestrian safety system, we design an efficient and real-time pedestrian distraction detection technique that overcomes some of these shortcomings. We demonstrate its practicality by implementing prototypes on commercially-available mobile and wearable devices and evaluating them using data collected from participants in realistic pedestrian experiments. Using these evaluations, we show that our technique achieves a favorable balance between computational efficiency, detection accuracy and energy consumption compared to some other techniques in the literature.
\end{abstract}

\begin{IEEEkeywords}
Pedestrian, Distraction, Hazard, Mobile, Wearables.
\end{IEEEkeywords}

\section{Introduction}
\label{intro}

Pedestrian safety has become a critical concern as the number of serious and fatal injuries due to pedestrian-related accidents continue to steadily rise every year \cite{williams2013pedestrian}. As one of the major causes of such pedestrian-related accidents, distracted driving has received significant attention over the past decade \cite{young2007driver,klauer2014distracted}, which has resulted in a host of techniques to detect and overcome distraction during driving. However, nearly 50\% \cite{bungum2005association} of all traffic related pedestrian deaths can be attributed to distraction among pedestrians (for example, inattentiveness while crossing roads and failure to obey traffic signs) rather than distracted drivers, which highlights the significant role pedestrian distraction plays in these accidents \cite{hyman2010did,schwebel2012distraction}. Besides this, distracted pedestrians are also susceptible to other non-traffic hazards in indoor and outdoor environments, such as falling over the edge of a subway platform, walking into obstacles, falling down a stairway, colliding with other pedestrians, and falling into an uncovered sewer manhole \cite{lang2013don}. It is evident that distracted pedestrians pose a significant threat not only to their own safety, but also to the safety of other pedestrians (and drivers), and effective systems and mechanisms to overcome these threats are critically needed. 

Designing effective pedestrian safety systems, however, has been challenging. A pedestrian safety system typically comprises two main components (Figure \ref{generic1}): (i) a \emph{distraction} or \emph{hazard detection} component, and (ii) an \emph{accident prevention} component. The advent of mobile and wearable devices (e.g., smartphones and smartwatches), equipped with a variety of high-precision sensors capable of capturing fine-grained user movements and context, provides a great opportunity to design sound distraction detection and recognition techniques. However, designing techniques that are accurate, efficient and real-time, is not straightforward, due to the memory, computation and communication limitations of these devices.

Several recent research efforts in the literature have attempted to improve pedestrian safety by detecting hazardous contexts (e.g., incoming vehicles, obstacles, uncovered manholes, etc.) with the help of data available from users' smartphone cameras \cite{wang2012walksafe,foerster2014spareeye,wei2015automatic,peng2010smartphone} or from specialized sensors (e.g., ultrasonic sensors or depth cameras) attached to the phones \cite{hincapie2013crashalert,ahn2013casual,wen2015we}. In addition to shortcomings such as reliance on smartphone camera feed or other specialized sensors and devices which limits their functionality, several of these schemes employ computationally-intensive data processing techniques that are challenging to implement on resource-constrained mobile and wearable devices. More importantly, the above techniques fail to generalize the problem of pedestrian distraction detection by not considering a diverse range of complex and concurrent activities that commonly resemble distraction, for example, detecting when users are walking, running or descending staircases and simultaneously reading, eating or drinking \cite{mwakalonge2015distracted,eatdistract}. As a result, the above solutions are unable to recognize a wide variety of distraction-related activities.

The key to designing a pedestrian safety system that has broad application and usage is to first generalize the problem of detecting distracted pedestrians as a \emph{concurrent activity recognition} (or \emph{CAR}) problem. Several robust and accurate CAR frameworks that detect and recognize a variety of human activities, and their complex combinations, by using data available from commercial mobile and wearable device sensors have already been proposed in the literature \cite{korpela2015energy,shoaib2016complex,liu2016complex,yan2012energy}. However, the applicability of these models for pervasive pedestrian distraction detection applications is unclear and has not been well-studied. It appears that a majority of these CAR models proposed in the literature, owing to their use of computationally expensive data processing and analysis techniques, could be challenging to implement and/or efficiently operate on consumer-grade mobile and wearable devices that possess limited computational and energy resources.

These shortcomings necessitate further investigation in two directions, which will be pursued by us in this paper: (i) \emph{is it possible to design a generic pedestrian distraction detection approach that can operate on existing commercial mobile and wearable devices and achieve a favorable balance between computational efficiency, detection accuracy, and energy consumption?} and (ii) \emph{how do existing concurrent activity recognition frameworks perform in a pedestrian distraction detection scenario?} In line with these objectives, we first  design a novel complex activity recognition technique, called \textit{Dominant Frequency-based Activity Matching (DFAM)}, which employs a lightweight frequency matching approach on motion (accelerometer and gyroscope) data available from users' mobile and wearable devices to accurately and efficiently detect and recognize a wide variety of complex distracted pedestrian related activities. Next, we undertake a comprehensive comparative evaluation of the proposed technique with well-known complex activity recognition approaches in the literature by means of distraction-related data collected from real human subject pedestrians.

\section{Background and Related Work}
\label{related}
We first outline significant mobile and/or wearable device based tools and techniques proposed in the literature for improving pedestrian safety, and discuss their limitations. As the pedestrian distraction detection problem can be generalized as a CAR problem, later we also discuss recent research results in the direction of concurrent activity recognition using these devices, primarily focusing on the recognition of human activities.

\subsection{Pedestrian Safety Systems}
Several research efforts in the literature have employed mobile and/or wearable devices to improve pedestrian safety by detecting hazardous contexts using users' smartphone camera \cite{wang2012walksafe,foerster2014spareeye,wei2015automatic,peng2010smartphone}.  \emph{WalkSafe} \cite{wang2012walksafe} utilized the rear camera of the smartphone to detect vehicles approaching a distracted user (or pedestrian) in order to promptly deliver a danger alert or notification. Deng et al. \cite{wei2015automatic} used image processing techniques and multi-sensor (barometer, accelerometer and gyroscope) information on smartphones to detect surrounding objects. Similarly, Peng et al. \cite{peng2010smartphone} used real time video processing of road traffic to help partially sighted pedestrians in spotting obstacles on their path. \emph{SpareEye} \cite{foerster2014spareeye} is another proposal which applied image processing techniques on a smartphone camera feed to simultaneously find multiple obstacles in a user's path unlike the results by Peng et al. \cite{peng2010smartphone}. One significant drawback of all these proposals is that they employ costly and resource-intensive image capture and processing techniques, which can adversely impact the performance and battery-life of mobile devices, thus diminishing their chances of being continually adopted by users. Reliance on a smartphone's camera also restricts the ability of these techniques to operate when the camera is obstructed, for example, in a user's pocket. 

Techniques for aiding pedestrian safety that do not rely on camera input, but rather on a smartphone's microphone \cite{lee2011acoustic} and GPS \cite{lin2016psafety} have also been proposed. For instance, Lee et al. \cite{lee2011acoustic} employ sound features extracted from the smartphone's microphone to detect oncoming vehicles, while \emph{pSafety} \cite{lin2016psafety} recognizes potential collisions between pedestrians and oncoming vehicles using the smartphone's GPS. One main shortcoming of these systems is that they are useful in detecting only outdoor traffic-related hazards scenarios. 

Furthermore, techniques that employ specialized devices and sensors for improving pedestrian safety have also been proposed. \emph{Lookup} \cite{jain2015lookup} uses information from specialized motion sensors attached to pedestrians' shoes to profile step and slope in order to detect curbs, ramps and other obstructions. Similarly, Ramos and Irani \cite{hincapie2013crashalert} used a depth camera (paired with a smartphone), while Ahn and Kim \cite{ahn2013casual} and \cite{wen2015we} employed an ultrasonic sensor for detecting pedestrian hazards and/or for guided navigation. Besides relying on specialized sensors or hardware, these systems attempt to address pedestrian safety by detecting obstacles or other potential hazards (to pedestrians). \emph{In this paper, we take an orthogonal approach to pedestrian safety by attempting to detect inattentiveness among pedestrians. Our main intuition for such an approach is that if distracted pedestrians are accurately detected in real-time, and promptly notified, they may be able to take appropriate corrective action to navigate away from a potentially hazardous situation}.

\subsection{Concurrent Activity Recognition (CAR)}
The problem of detecting distracted pedestrians can be generalized as a concurrent activity recognition or CAR problem where the goal is to detect concurrent pedestrian activities of being mobile (e.g., walking, running or climbing/descending stairs) and being distracted (e.g., texting, eating or reading). CAR techniques that can distinguish different combinations of elementary activities have been extensively used in the literature for complex human activity recognition. For instance, Shoaib et al. \cite{shoaib2016complex,shoaib2016hierarchical} used \textit{multi-source} and \textit{multi-sensor motion} data, from two smartphones, one in trouser pocket and the other on the wrist, to recognize activities that involve hand gestures, such as smoking, eating, drinking coffee and giving a talk. Liu et al. \cite{liu2016complex} also employed multi-sensor time series data to recognize sequential, concurrent, and generic complex activities by building a dictionary of the time series patterns (called \emph{shapelets}) to represent atomic activities. 

However several shortcomings in these approaches, as outlined below, prevent them from being effectively used in pedestrian safety applications. For instance, Shoaib et al.'s work \cite{shoaib2016hierarchical} requires the system to keep track of time segments that precede and follow the current one, and thus, unsuitable for pedestrian safety applications that require real-time operation and feedback. Others are not suitable for implementation on resource-constrained mobile and wearable devices, primarily due to their use of complex feature sets and classification functions. As discussed before, one of the main functional requirement for a mobile/wearable device based CAR framework for pedestrian safety is computational and energy efficiency. Earlier research efforts in energy-aware recognition mechanisms \cite{yan2012energy} have achieved a favorable balance between classification accuracy and energy consumption, but these schemes have been successful in recognizing only simple activities, such as standing, walking and sitting, but not concurrent activities. Recently, Korpela et al. \cite{korpela2015energy} proposed an energy-aware CAR framework for real-time applications by using a minimal feature set to recognize individual data segments and a hierarchical classification mechanism for concurrent activity recognition. However, their framework employs a specialized wearable device hardware, and may not work for commercial off-the-shelf mobile devices thus making them less likely to be adopted by users.

\subsection{A generic pedestrian safety framework}
\begin{figure}[t]
	\includegraphics[width=\linewidth]{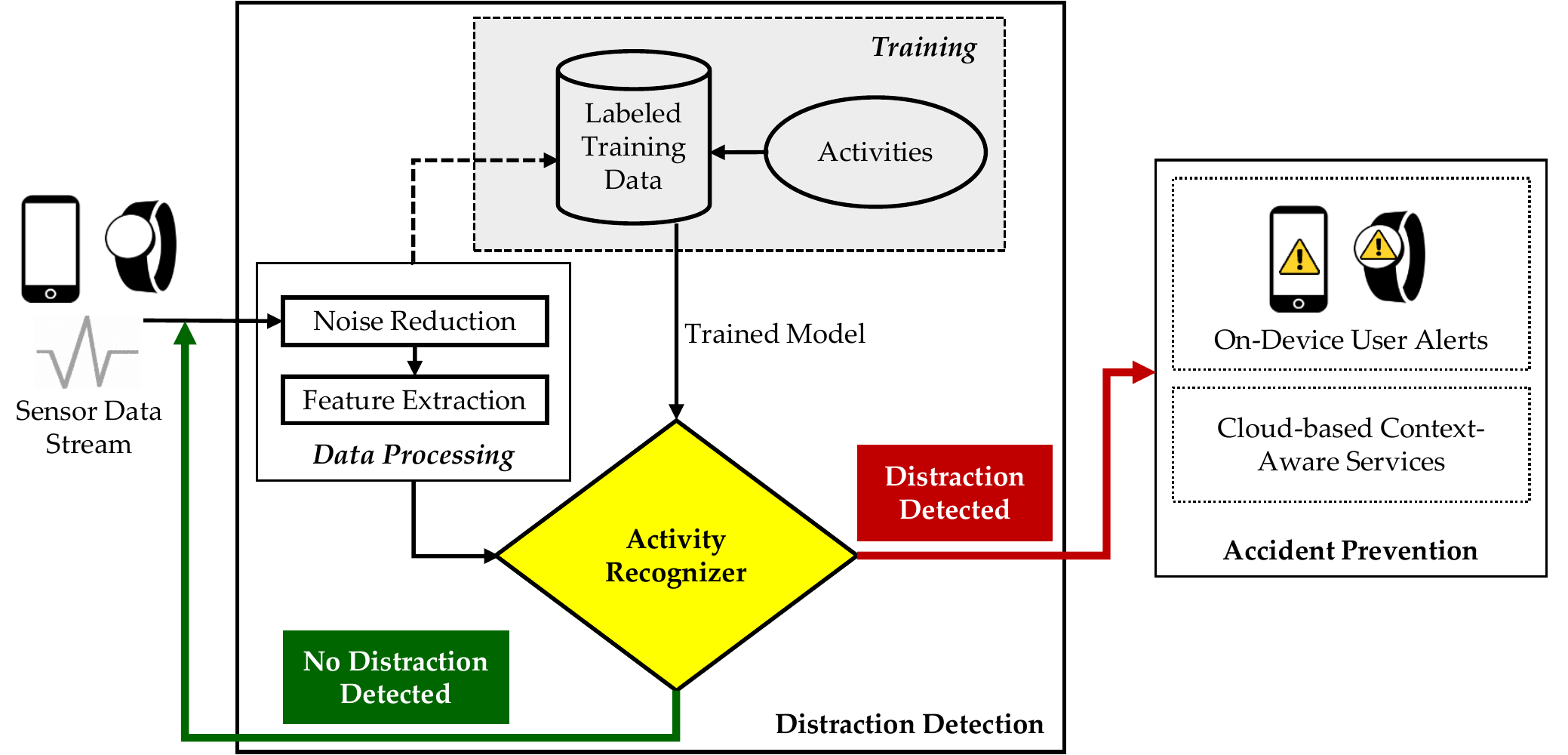}
	\caption{A generic pedestrian safety system.}
	\label{generic1}
\end{figure}

The two main components of a pedestrian safety system include (Figure \ref{generic1}): (i) a \emph{distraction} or \emph{hazard detection} component, and (ii) an \emph{accident prevention} component. In this paper, we primarily focus on the former. Figure \ref{generic1} depicts the design of a generalized learning-based framework which is the main building block for pedestrian distraction detection in such systems. As shown in Figure \ref{generic1}, the distraction detection framework comprises of: (i) a \emph{data processing module} (includes noise removal, segmentation, and feature generation processes), and (ii) a \emph{CAR model building phase} (includes the design of an appropriate activity classification function and training it using processed labeled training data). Once a trained CAR model is available, it can be used to recognize (or classify) distracted pedestrian activities. Such a design of the distraction detection framework is commonly employed in the literature (and in practice) for pedestrian safety and other applications, and will also be employed by us in this paper. 

Our distraction detection framework relies on multi-sensor data obtainable from multiple mobile devices carried by the pedestrians, specifically, motion (including, data from accelerometer and gyroscope sensors) and contextual information from the pedestrian's smartphone and smartwatch. The data processing module in our framework filters this multi-sensor data to eliminate errors and inconsistencies, segments it into fixed-size blocks or windows, and extracts relevant features from it. It should, however, be noted that the data processing task may vary depending on the chosen CAR technique. The extracted features are then used to train appropriate CAR models within a supervised learning paradigm. These trained CAR models are utilized by our framework for distracted activity classification or recognition tasks. It should be easy to see that an appropriate CAR technique is central to the design of a pedestrian distraction detection framework that can attain a practical balance between computational efficiency, detection accuracy and energy consumption, and is the focus of our research in this paper.

In our preliminary work \cite{vinayaga2017towards} towards designing an efficient CAR model suitable for detecting distracted pedestrian activities, we proposed a novel CAR technique called DFAM. We also undertook a preliminary evaluation of DFAM primarily in a personalized setting, where both training and testing of DFAM were conducted using the same participants' motion data. In this paper, we significantly expand our DFAM evaluation in a personalized setting, and comprehensively evaluate a generalized setting where the training and testing are conducted on disjoint sets of participants. Such a generalized evaluation can validate if DFAM can accurately detect distracted pedestrian activities even when personalized training data from a user is not available. We were also able to significantly reduce the resource footprint of our initial DFAM implementation by designing a hierarchical activity recognition model, presented and evaluated in Section \ref{idfam}.

\section{Proposed Pedestrian Distraction Detection Technique}
\label{system}
A majority of the time domain features utilized for concurrent activity recognition \cite{shoaib2016complex, shoaib2016hierarchical} are computationally intensive and thus not suitable for real-time pedestrian distraction detection. Moreover, with multi-source (smartphone and smartwatch) data, data fusion (before feature extraction) proves to be a challenging task owing to source (time) synchronization issues in high-precision data streams. \emph{Our proposed CAR technique, referred to as DFAM, addresses this problem, and is capable of computing features (frequency domain, particularly dominant frequencies) directly and independently on the different devices, such as smartphone and smartwatch}. Technical details of our proposed CAR technique are presented in this section. We also outline other well-known techniques that have been employed in the literature \cite{shoaib2016complex,shoaib2016hierarchical} for similar activity classification tasks, as we later empirically compare the performance of our DFAM technique against these classical activity classification techniques. 

\subsection{Dominant Frequency-based Activity Matching (DFAM)} 
 DFAM is inspired by the audio matching algorithm proposed by Avery Wang \cite{wang2003industrial}. Proprietary versions of Wang's algorithm are commonly used in popular song searching applications, such as \emph{Shazam}. Due to the significant differences between audio (found in audio files) and motion data (sampled from the smartphone and smartwatches), it is non-trivial to use Wang's audio matching algorithm directly for activity recognition using motion data. In the audio matching application, matching features in the test audio file occur at almost identical relative time offsets from the beginning of the audio file being matched to. In contrast, motion data from pedestrian activities generally does not occur at exactly fixed time offsets, therefore requiring a new matching algorithm. Other differences between motion and audio data include a significantly lower sampling rate of smartphone and smartwatch motion sensors (compared to audio data which is generally sampled at a much higher frequency) and distinctly different \emph{dominant frequency} ranges of both types of data. 
Recently, Sharma et al. \cite{sharma2008frequency} successfully applied \emph{dominant frequency-based activity matching} for simple (non-concurrent) activities, using fixed threshold-based classifiers. In this paper, we use preprocessing techniques used by Wang \cite{wang2003industrial} and extend Sharma et al.'s work significantly in order to recognize concurrent activities related to pedestrian distractions.

DFAM utilizes multiple \emph{frequency bins} to extract more than one dominant frequency, which can be used to characterize and recognize activities more accurately. As a result, by using frequency domain features, we can avoid resource intensive sample-to-sample synchronization required before computing multi-source time domain features. Our proposed DFAM scheme comprises of a model generation or training phase and a classification phase, as outlined next.

\subsubsection{Model Generation}
\label{dft}
During the training phase, (low-pass) filtered time-series motion data from the smartphone and smartwatch, denoted as $T_p$ and $T_w$, respectively, corresponding to each activity of interest is first \emph{segmented} into smaller fixed-sized windows of $W$ samples in a sliding fashion with possible overlap for maximum data utilization. Let's assume that this motion data is sampled at a frequency $f_s$. 

{\scriptsize
\begin{equation}
T_p = \{\textsuperscript{1}b_p, \textsuperscript{2}b_p, \ldots, \textsuperscript{m}b_p\} \textrm{;} \quad T_w = \{\textsuperscript{1}b_w, \textsuperscript{2}b_w, \ldots, \textsuperscript{n}b_w\}
\end{equation}
}
{\scriptsize
\begin{equation*}
\textrm{where} \quad m = \frac{sizeof(T_p)}{W} \quad \textrm{and} \quad n = \frac{sizeof(T_w)}{W}
\end{equation*}
}
After this pre-processing step, the frequency response of each window in $T_p$ and $T_w$ is independently calculated using a discrete Fourier transformation technique such as a fast Fourier transform (or FFT \cite{batenkov2005fast}). Let the frequency responses corresponding to $T_p$ and $T_w$ be represented as $F_p$ and $F_w$, respectively.

{\scriptsize
\begin{equation}
F_p = \{\textsuperscript{1}s_p, \textsuperscript{2}s_p, \ldots, \textsuperscript{m}s_p\} \textrm{;} \quad F_w = \{\textsuperscript{1}s_w, \textsuperscript{2}s_w, \ldots, \textsuperscript{n}s_w\}
\end{equation}
}
{\scriptsize
\begin{equation*}
\textrm{where} \quad \textsuperscript{i}s_p = FFT(\textsuperscript{i}b_p) \quad \textrm{and} \quad \textsuperscript{i}s_w = FFT(\textsuperscript{i}b_w)
\end{equation*}
}
Frequency response of all windows, $\textsuperscript{i}s_p \in F_p$ and $\textsuperscript{i}s_w \in F_w$, are then analyzed for a dominant frequency (DF) in $g$ frequency bins, with one dominant frequency in each bin:

{\scriptsize
\begin{equation*}
DF: \{f_{(0,u_1)}, f_{(u_1,u_2)}, \dots, f_{(u_{g-1},\frac{f_s}{2})}\} 
\end{equation*}
} 
All of the observed dominant frequency in each of the $g$ bins are then mapped to one of the $h$ buckets and combined to create a `\emph{signature}' for the activity. The buckets are obtained by splitting the entire frequency spectrum (or part of the spectrum covered by each bin) into $h$ equally sized intervals $\small [{f_o, f_c}]$ with constant \small$l = f_c - f_o$\normalsize. 

{\scriptsize
\begin{equation}
\quad H \colon v \to \lfloor v/_l \rceil \bmod h \textrm{,} \quad H \colon \mathbb{R}\textsuperscript{g} \to \mathbb{Z}\textsuperscript{g}_{< h} 
\end{equation}
}
{\scriptsize
\begin{equation}
H_x = H(DF_{x}); \quad H_y= H(DF_{y}); \quad H_z = H(DF_{z})
\end{equation}
}
The hash functions \small $H6-H8$ \normalsize from Table \ref{tab:functions} are dependant on $W$ as $h$ is equivalent to \small $W, \lfloor\frac{W}{2}\rceil, \lfloor\frac{W}{3}\rceil$ \normalsize respectively. The remaining functions \small$H0-H5$ \normalsize have bucket size $h$ independent of window size $W$.
\begin{table}[h]
  \centering
  \caption{Summary of hash functions used.}
  \scriptsize
    \begin{tabular}{|c|l|}
    \toprule
    Function & Definition \\
    \midrule
    $H0$    &  $\quad v \to \lfloor 8g(\lfloor v \rceil \bmod \lceil \frac{f_s}{2g} \rceil) / f_s\rfloor$ \\
    $H1$    &  $\quad  v \to \lfloor g(\lfloor v \rceil \bmod \lceil \frac{f_s}{2g}\rceil) / \frac{f_s}{2g}\rfloor$ \\
    $H2$    &  $\quad  v \to \lfloor 6g(\lfloor v \rceil \bmod \lceil \frac{f_s}{2g} \rceil) / f_s\rfloor$ \\
    $H3$    &  $\quad  v \to \lfloor 4g(\lfloor v \rceil \bmod \lceil \frac{f_s}{2g} \rceil) / f_s\rfloor$ \\
    $H4$     &  $\quad  v \to \lfloor v/_2 \rceil \bmod \lfloor\frac{f_s}{4}\rceil$ \\
    $H5$     &  $\quad  v \to \lfloor v \rceil \bmod \lfloor\frac{f_s}{2}\rceil$ \\
    $H6$     &  $\quad  v \to \lfloor 2vW/_{f_s}\rceil \bmod W$ \\
    $H7$	&  $\quad  v \to \lfloor vW/_{f_s} \rceil \bmod \lfloor\frac{W}{2}\rceil$ \\
    $H8$     &  $\quad  v \to \lfloor 2vW/_3{f_s} \rceil \bmod \lfloor\frac{W}{3}\rceil$ \\
    \bottomrule
    \end{tabular}%
  \label{tab:functions}%
\end{table}%

As we are employing multiple devices and sensors, with each sensor possibly outputting measurements across multiple dimensions (e.g., each accelerometer sensor measurement is across three dimensions), each training data point will consist of measurements across multiple dimensions. For example, a dominant frequency analysis on three-dimensional ($x, y, z$) time series data window will result in a three-dimensional training point $\langle H_x, H_y, H_z\rangle$, where $H_x$, $H_y$, and $H_z$ are the hashes of dominant frequencies on respective axes. Now, let us denote the set of all distracted activities as $\mathbb{D}$, and the set of all pedestrian activities as $\mathbb{P}$. For each activity $a_u \in\mathbb{P}$, a training dataset made of equalized data points is created during the training phase, and stored along with the corresponding label $a_u$. Similarly, for each concurrent activity $a_v \in\mathbb{P}\times\mathbb{D}$, another training dataset made of equalized data points is created during the training phase, and stored along with the corresponding label $a_{v}$.

\subsubsection{Activity Classification}
To correctly classify a (test) user activity (say, $a_{c}$), DFAM employs a dominant frequency matching technique using the labeled training data (obtained from the model generation phase), as described below. Given a test window with $s$-axis signatures, the activity is matched using the following scoring function:

{\scriptsize
\[ S_{i,j}(a_{c}) =
  \begin{cases}
    0       & \quad \text{if } \sum_{k=1}^{s} F(\textsuperscript{c}H_k , \textsuperscript{train}H^{i,j}_k) = 0\\
    (\frac{1}{s})^s  & \quad \text{if } \sum_{k=1}^{s} F(\textsuperscript{c}H_k , \textsuperscript{train}H^{i,j}_k) = 1\\
    (\frac{2}{s})^s  & \quad \text{if } \sum_{k=1}^{s} F(\textsuperscript{c}H_k , \textsuperscript{train}H^{i,j}_k) = 2\\
    \quad \vdots  & \quad \quad \quad \quad \quad \quad \quad \vdots\\
    (\frac{s-1}{s})^s  & \quad \text{if } \sum_{k=1}^{s} F(\textsuperscript{c}H_k , \textsuperscript{train}H^{i,j}_k) = s-1\\
    1  & \quad \text{if } \sum_{k=1}^{s} F(\textsuperscript{c}H_k , \textsuperscript{train}H^{i,j}_k = s\\
  \end{cases}
\]
}

where $S_{i,j}(a_{c})$ is the matching score per training instance $j$ in each activity $a_i\in\mathbb{P}\times\mathbb{D}$, $\textsuperscript{c}H_k$ is the current activity signature from $k$-th sensor axis, $\textsuperscript{train}H^{i,j}_k$ is the signature from $k$-th sensor axis of $j$-th training instance of activity $a_i$, and 

{\scriptsize
\[ F(a,b) =
  \begin{cases}
    0 & a \neq b\\
    1 & a = b\\
  \end{cases}
\]
}

The above scoring function gives exponentially more weight to multi-dimensional signature matches, which will intuitively result in a higher score when matching with the ground truth activity. Finally, the activity is classified after matching against the entire training dataset of all activities as follows:

{\scriptsize
\begin{equation}
\\ arg\max_{i}  \sum\limits_{j} S_{i,j}(a_{c}) \quad \forall a_i \in \mathbb{P}\times\mathbb{D}
\label{egclass}
\end{equation}
}

The current activity $a_{c}$ is then classified as that activity $a_i$ which achieves the maximum aggregated score as shown in Equation \ref{egclass}. A list of symbols used in the above system model is summarized in Table \ref{tab:symbols}.

\begin{table}[htbp]
  \centering
  \caption{Summary of symbols used.}
  \scriptsize
    \begin{tabular}{|c|p{6.5cm}|}
    \toprule
    Symbol & Definition \\
    \midrule
    $T_p$    & Time-series motion data from smartphone \\
    $T_w$    & Time-series motion data from smartwatch \\
    $F_p$    & Frequency responses corresponding to $T_p$ \\
    $F_w$    & Frequency responses corresponding to $T_w$ \\
    $W$     & Number of samples in a training and testing window \\
    $r$     & Percentage of window overlap when extracting training windows from $T_p$ and $T_w$ \\
    $g$     & Number of frequency bins \\
    $h$	& Number of buckets associated with the hash function H \\
    $\mathbb{P}$     & Set of pedestrian activities \\
    $\mathbb{D}$     & Set of distracted activities \\
    \bottomrule
    \end{tabular}%
  \label{tab:symbols}%
\end{table}%

\subsection{Classical Classification Models}
Traditional supervised learning-based classification functions, such as \emph{Naive Bayes (NB)}, \emph{Decision Tree (DT)}, \emph{Random Forests (RF)}, \emph{Support Vector Machine (SVM)}, \emph{k-Nearest Neighbours (k-NN)}, have been successfully used in the literature (and in several deployed commercial applications) for detecting complex and concurrent human activities \cite{shoaib2016complex, shoaib2016hierarchical}. Given that distracted pedestrian activities are inherently concurrent activities, these supervised learning-based techniques comprise of a suitable candidate set for a comparative performance evaluation with our proposed DFAM technique. Below, we outline how these classification techniques are employed within our pedestrian distraction detection framework, and provide details on the related data pre-processing, feature extraction and model training tasks.   

Firstly, the (low-pass) filtered time-series motion data from the smartphone and smartwatch, denoted as $T_p$ and $T_w$ respectively, corresponding to each activity of interest is \emph{segmented} into smaller fixed-sized windows, as discussed earlier for DFAM. Each of the motion data stream $T_p$ and $T_w$ comprises of both the accelerometer and gyroscope sensor data sampled along all the three axes at some frequency $f_s$. A set of time and frequency domain features, as have been employed in the literature \cite{shoaib2016complex,sun2010activity,varkey2012human,parate2014risq,ilmjarv2015detecting} for activity recognition (and briefly outlined below), are then computed from each window of the time series motion data streams:
\begin{itemize}[leftmargin=*]
\item Mean, minimum, maximum, standard deviation, variance, along with energy and entropy of discrete FFT components for each of the three axes of both the accelerometer and gyroscope time-series data.  
\item Root mean square (RMS) correlation measures among the three axes for each of the accelerometer and gyroscope data.
\item Mean, median, and maximum of the instantaneous speed (only for the accelerometer data).
\item Mean, median, and maximum of roll velocity (only for the gyroscope data).
\end{itemize}  

During the training phase, the motion time-series of each activity is uniformly segmented into training windows of $W$ samples, and the above features are computed for all available training windows. The entire process (filtering, segmentation, and feature extraction) is repeated for all the considered distraction-related activities in $\mathbb{D}$ and non-distraction activities in $\mathbb{P}$ in the training dataset to create a labeled feature set for all the activities. Such a labeled training (feature) set is then used to train each of the concurrent activity classification models. These data pre-processing and feature extraction tasks remain the same for all the supervised classification functions. We describe some commonly employed classification models in the literature below:

\subsubsection{Naive Bayes (NB)}
Given a test (or unlabeled) feature set, a trained NB model estimates the posterior probabilities of each activity (in the set of all considered concurrent activities $\mathbb{P}\times\mathbb{D}$), assuming that the input features are independent of each other. The unknown or test activity is then assigned an activity label corresponding to the maximum posterior probability value. Posterior probabilities are estimated according to the Bayes rule.
%

\subsubsection{Decision or Classification Tree (DT)} 
This technique constructs a tree structure using the labeled feature sets of the training data, where leaves of the tree represent the class labels of the different concurrent activities, whereas the branches represent conjunction of features that lead to these class labels. Now given the feature set of an unknown activity, the corresponding activity label is determined by traversing through the branches of the trained tree model using the discrete feature values in the feature set until a leaf node is reached. The unknown activity is then classified with the label corresponding to the reached leaf node.

\subsubsection{Random Forests (RF)} 
These are ensembles of decision trees that could output multiple activity labels, one each from a decision tree, for an unknown activity sample. The unknown activity is then assigned an activity label using a majority rule.

\subsubsection{Support Vector Machine (SVM)} 
This technique uses the labeled training data (feature sets) to learn the hyperplanes separating the different activity classes. These hyperplanes or decision boundaries are optimized to achieve the maximum separation distance between activity classes. After the model is trained, i.e. separating hyperplanes are determined, the feature set corresponding to an unknown activity is assigned a label corresponding to its placement in the bounded $n$-dimensional feature space.

\subsubsection{k-Nearest Neighbours (k-NN)} 
This technique creates a trained classification model by grouping labeled training data or feature sets into separate clusters based on their class label or activity they represent. Then, a data sample from an unknown activity (or an unlabeled feature set) is classified as the class (or activity) of a majority of its $k$ closest or nearest neighbors. Euclidean distance can be used as a closeness measure to compute the proximity between two feature sets.

\section{Performance Evaluation}
\label{evaluation}

\subsection{Experimental Setup}
\label{placeset}
\begin{figure}[]
\centering
\includegraphics[width=\linewidth]{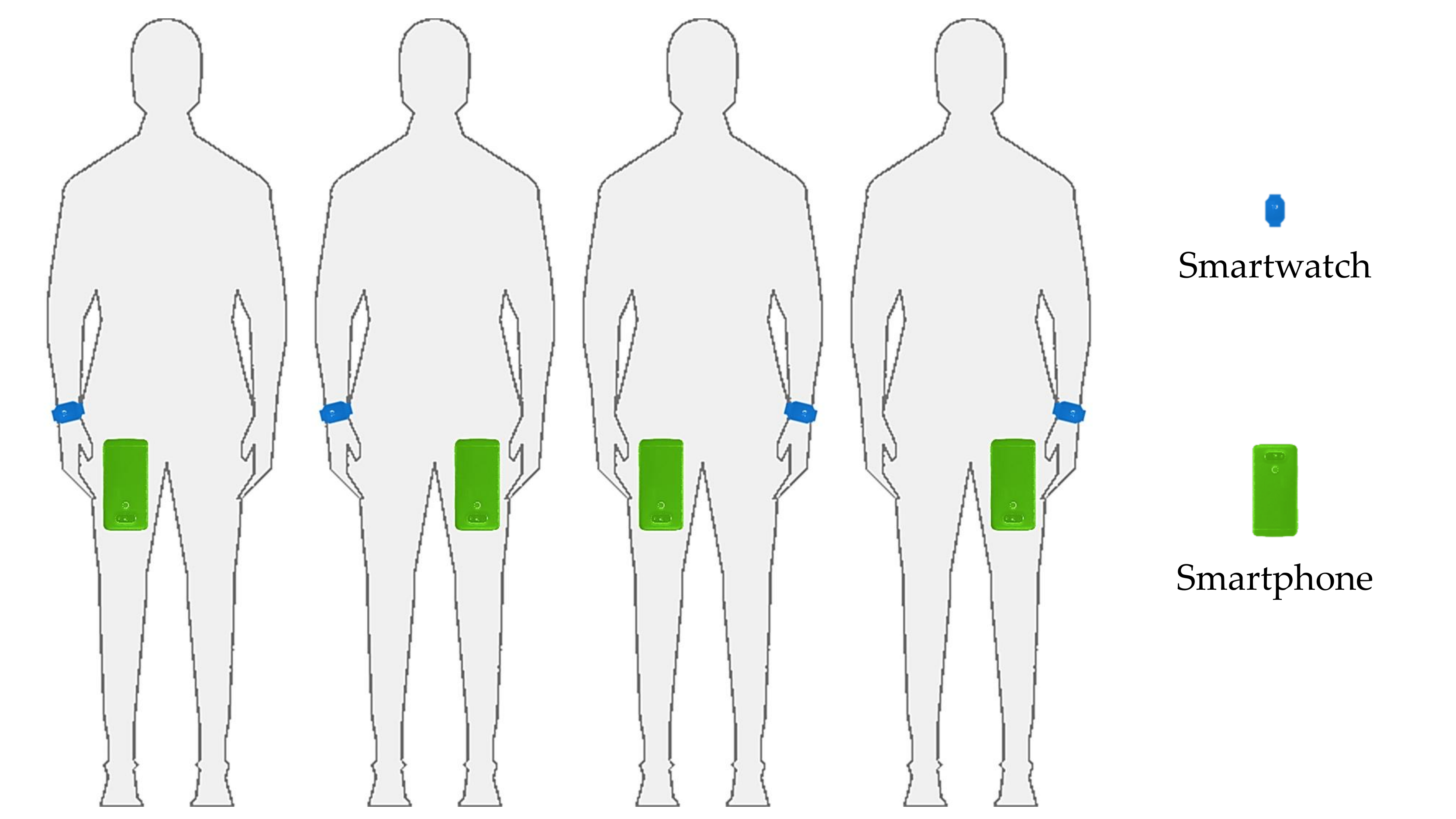}
\caption{Smartwatch and smartphone placements during data collection (RR, RL, LR, LL).}
\label{positions}
\end{figure}

Our experiments, approved by Wichita State University's Institutional Review Board (IRB), involved collection of motion sensor data from twenty participants. Participants were primarily from Wichita State University between the age groups of 20 and 40, among which 15 were male and the remaining female. Each experiment belonged to one of the four device-placement scenarios (Figure \ref{positions}): two same side, two different side device placements. In the same-side placement scenarios, either both smartwatch and smartphone are worn on the right wrist and placed inside right hip pocket (RR), or worn on the left wrist and placed inside the left hip pocket (LL). The latter two scenarios alternate the placements to the opposite sides of the body, i.e., smartwatch on right wrist along with phone in left hip pocket (RL), and smartwatch on the left wrist with the phone in right pocket (LR).

\subsubsection{Data Collection}
Each participant in our experimental setup was assigned a Motorola Moto XT1096 smartphone along with a paired smartwatch (Sony Smartwatch 3 or a LG Urbane W150) in order to record motion sensor data for a predefined set of pedestrian activities as outlined in Appendix A. During the enaction of these activities, the wrist-worn smartwatch and paired smartphone recorded the three-dimensional accelerometer and gyroscope sensor data corresponding to these activities at a sampling frequency of 50$Hz$. The activities performed by the participants encompassed simple pedestrian activities, and concurrent non-distracted and distracted pedestrian activities. All concurrent activities except the starred (*) activities in Table \ref{activitylist} constitute a set of distracted pedestrian activities. Their execution sequence was randomized across participants in order to minimize any order effect in the experiments. Participants were compensated with a maximum of \$40 or 8 SONA credits\footnote{Wichita State University's psychology SONA research credits, where participating students received compensation in the form of research credits as a course requirement.} for their participation. Table \ref{datasetlist} outlines the number of datasets collected for each device-placement scenario.

\begin{table}[h]
\scriptsize
\centering
\caption{Activities performed by the participants.}
\label{activitylist}
    \begin{tabular}{|p{2.0cm}|p{4.0cm}|}
    \toprule
    Simple Activities & Concurrent Activities \\
    \midrule
    Standing  & Walking + Using Smartphone \\
    Walking & Walking + Reading \\
    Climbing stairs & Walking + Eating \\
    Descending stairs & Walking + Drinking \\
    Sitting & Climbing stairs + Using Smartphone \\
    Running & Climbing stairs + Reading \\
    & Climbing stairs + Eating \\
    & Climbing stairs + Drinking \\
    & Descending stairs + Using Smartphone\\
    & Descending stairs + Eating \\
    & Descending stairs + Reading\\
    & Descending stairs + Drinking \\
    & Running + Using Smartphone  \\
    & Standing + Reading*  \\
    & Standing + Eating*  \\
    & Standing + Drinking*  \\
    & Sitting + Using Smartphone* \\
    & Standing + Using Smartphone*  \\
    \bottomrule
    \end{tabular}
\end{table}

\begin{table}[h]
  \centering
  \scriptsize
  \caption{Datasets per placement scenario.}
  \label{datasetlist}
    \begin{tabular}{|r|rrrr|r|}
    \toprule
     & RR & LL & RL & LR & Total \\
    \midrule
    LG+Moto & 5     & 5     & 5     & 5     & 20 \\
    Sony+Moto & 5     & 5     & 5     & 5     & 20 \\
    \midrule
    Total & 10    & 10    & 10     & 10     & 40 \\
    \bottomrule
    \end{tabular}
\end{table}

Collecting data from real distracted pedestrians to evaluate the proposed detection mechanisms in a safe, yet realistic setting, has been one of the biggest challenges of this research effort. To ensure participant safety, we conducted all data collection in a controlled, but realistic, environment with some activities conducted inside a building, while some others in the open space. Ensuring participant safety during certain distracted activities (e.g., descending stairs while reading) was challenging due to potential falling and injury risks. In order to overcome these challenges, we took several precautionary measures as per the IRB's recommendation throughout the data collection. For example, a safety harness was placed on the participant when climbing up and down the stairs and performing a distracted activity such as reading or eating. The safety harness was handled by two researchers while the participant was performing the activity, and was designed so as to not impede the participants' natural motion. Participants were also allowed sufficient rest and water breaks between activities to avoid injuries due to fatigue and dehydration. Due to these safety precautions, our data collection experiments took a significant amount of time -- each participant took, on an average, 2-3 hours to complete all the activities. Also, due to the intensive physical demands of the (data collection) experiments, participants had to be filtered based on their fitness level (as required by the IRB). The physical demands of our experiments, together with these additional constraints in selecting participants limited our ability to recruit an even large number of participants. Despite the above challenges, we feel that we have collected an adequate amount of data to make statistically significant validations. Lastly, activities such as eating, drinking or reading while running were not included during the data collection experiments as per IRB's recommendation to ensure participant safety.

\subsubsection{Evaluation Parameters}
Analysis of the proposed DFAM technique was done on a 64-bit PC with Intel Xeon processor and 32 GB RAM using Java 8 and Python 3. This PC implementation was used to extensively benchmark performance, which is not feasible on a resource constrained smartphone. We also implemented DFAM for the Android and Android Wear ecosystem, which was used to evaluate on-device response time and resource utilization under real-life usage. For CAR models deployed on the Motorola Moto XT1096 smartphone, the following metrics captured system responsiveness and resource efficiency. Response time (RT) is the time taken by the system to read activity data, process the motion sensor data block, generate features, classify the activity, and notify the user. It is measured in seconds ($s$). 
CPU utilization is measured as a percentage of total CPU resources utilized by the application. RAM usage denotes the amount of total RAM used by the application, and is measured in megabytes (MB). Battery usage indicates the percentage of the total battery capacity used by the application, measured in milliwatts ($mW$). The CPU utilization, battery usage and RAM utilization were recorded in these trials over a period of sixty minutes and repeated ten times to validate the results. 

\subsection{DFAM Performance} 
\label{dfamperformance}
First, we evaluate the classification performance of our proposed DFAM technique under varying model parameters such as window size, hash function, bin size and device placement scenarios, in a personalized setting and using the combined Sony+Moto datasets (as outlined earlier in Table \ref{datasetlist}). In a personalized evaluation setting, a DFAM model is trained with data belonging to only one participant, and validated with the same participant's data. Each participant dataset was trained and analyzed individually and independently of the other datasets. After training, each model is evaluated using a $10$-fold cross validation, where the dataset is randomly split into ten equal parts, one of which becomes the test set and the remaining nine parts constitute as the training set. The testing is repeated on each of the ten parts as the test set, and results are averaged across the ten tests.

\subsubsection{Training Window Overlap Ratio} 
\label{tor}
Training window overlap ratio ($r$) is the fraction of common samples between two consecutive time-series windows selected from the entire training time-series data ($T_p$ and $T_w$). We conduct an empirical analysis to better understand how a trained DFAM model will be affected by different training window overlap ratios, and determine the optimal value of $r$ for the purpose of detecting distracted pedestrian related activities. The training window overlap ratio is also correlated to the number of training windows that can be extracted from the training data, which has significant implications if training is conducted in a personalized setting. In any personalized activity recognition model, users are supposed to train the model by providing it with representative training data for each type of activity to be detected. However, if the amount of personalized data required for training is large, the training process may become long and tiresome for the user. Therefore, if the model performs satisfactorily for high training window overlap ratios, it would imply that DFAM will require less training data, and thus can have a concise and user-friendly training process. 

\begin{figure}[]
\centering
\includegraphics[width=\linewidth]{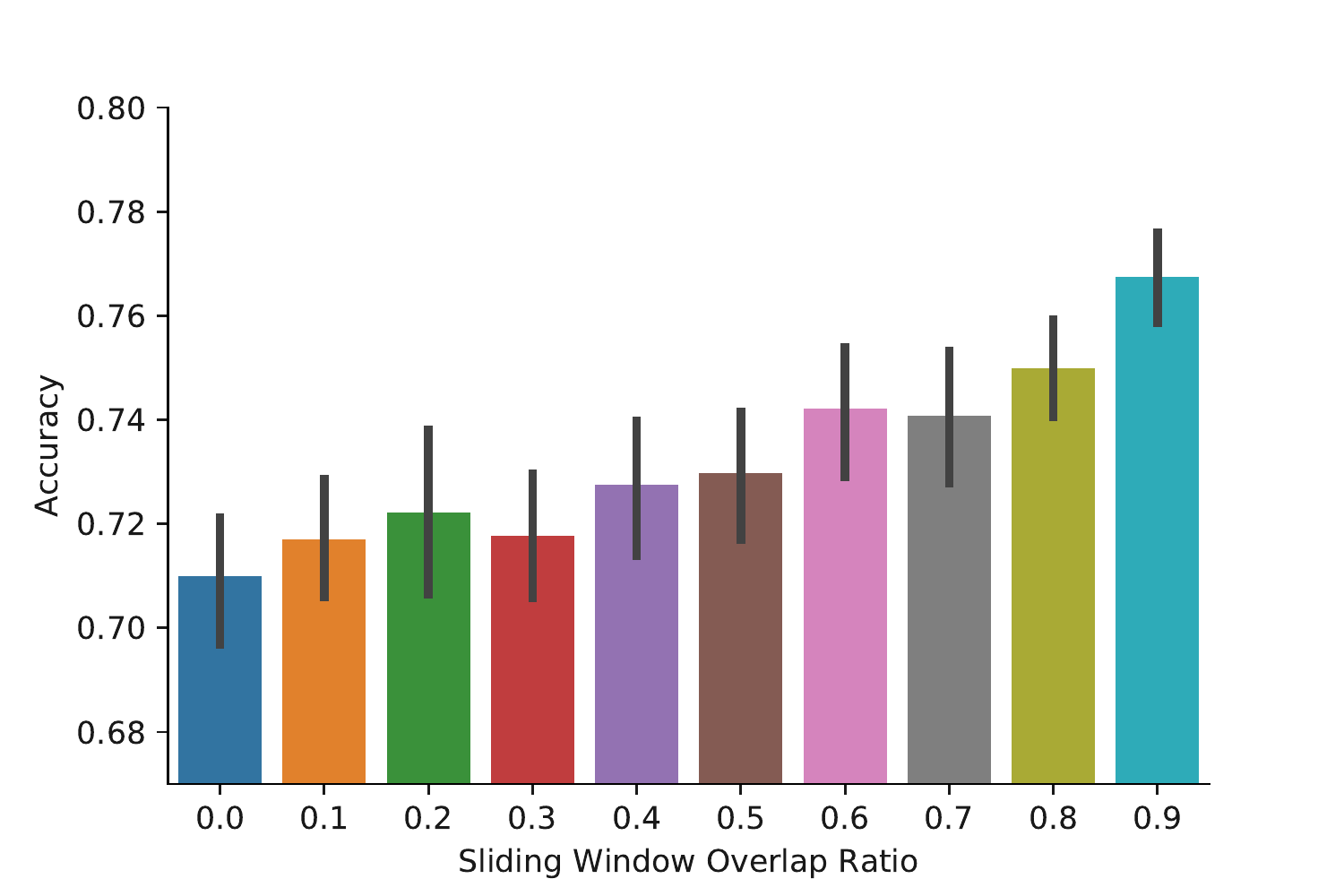}
\caption{Personalized DFAM classification accuracy for different window overlap ratios. $W=128$, $g=3$.}
\label{trainoverlap}
\end{figure}

Figure \ref{trainoverlap} shows an example of how DFAM classification accuracy changes for different values of $r$, obtained after a single participant trained the model for personalized use. To avoid inconsistencies in this visualized example, the same training time-series data ($T_p$ and $T_w$) were used for all values of $r$. For the sake of simplicity, we also fix other critical parameters that can also affect DFAM classification accuracies (evaluated in later sections), such as window size ($W=128$), number of frequency bins ($g=3$), and hash function ($H7$). In Figure \ref{trainoverlap}, we can observe that the trained models generated using lower $r$ values resulted in lower accuracy, and vice versa. For instance, the accuracy for DFAM model trained with $r=0.1$ is 0.72, whereas DFAM trained with $r=0.9$ resulted in the accuracy of 0.77. \emph{As a result, DFAM can perform satisfactorily with less training data, thus requiring a concise training process.} 

However, for a DFAM with a window of 32 samples, and a training overlap ratio $r=0.9$, a maximum of 927 signatures can be extracted for each activity time-series data worth 60 seconds, which is much larger than a DFAM with $r=0.7$ and approximately 309 signatures. If the DFAM is trained with data from 24 activities, the signature count increases drastically, for the $r=0.9$ scenario, to 22248 when compared to the $r=0.7$ scenario which has 7416 signatures. Therefore, for higher $r$ values, DFAM has more signatures in order to match against to recognize an activity, which may increase response time. Considering the trade-off between accuracy and response time, we select a middle ground, $r=0.7$, for analysis of other parameters affecting DFAM, in the following subsections.



\subsubsection{Hash Function} 
The hash function in DFAM (Table \ref{positions}), helps to reduce the final model size and response time. This is mainly because it takes lesser space to store and lesser time to match against a mapped value of a dominant frequency, say in two characters, than in its original floating format. Different hash functions generate dissimilar signatures for the same set of dominant frequencies, which may affect the pedestrian activity detection accuracy. This section investigates how the performance of DFAM is affected by various hash functions based on a fixed frequency range 0-25$Hz$, which is determined by the 50$Hz$ sensor sampling rate. The functions are precursors to generating DFAM model representations that consume fewer resources in terms of space and processing time. Their adoption is likely to provide extra service time, which may have a positive impact on the overall user experience.


\begin{figure}[]
\centering
\includegraphics[width=\linewidth]{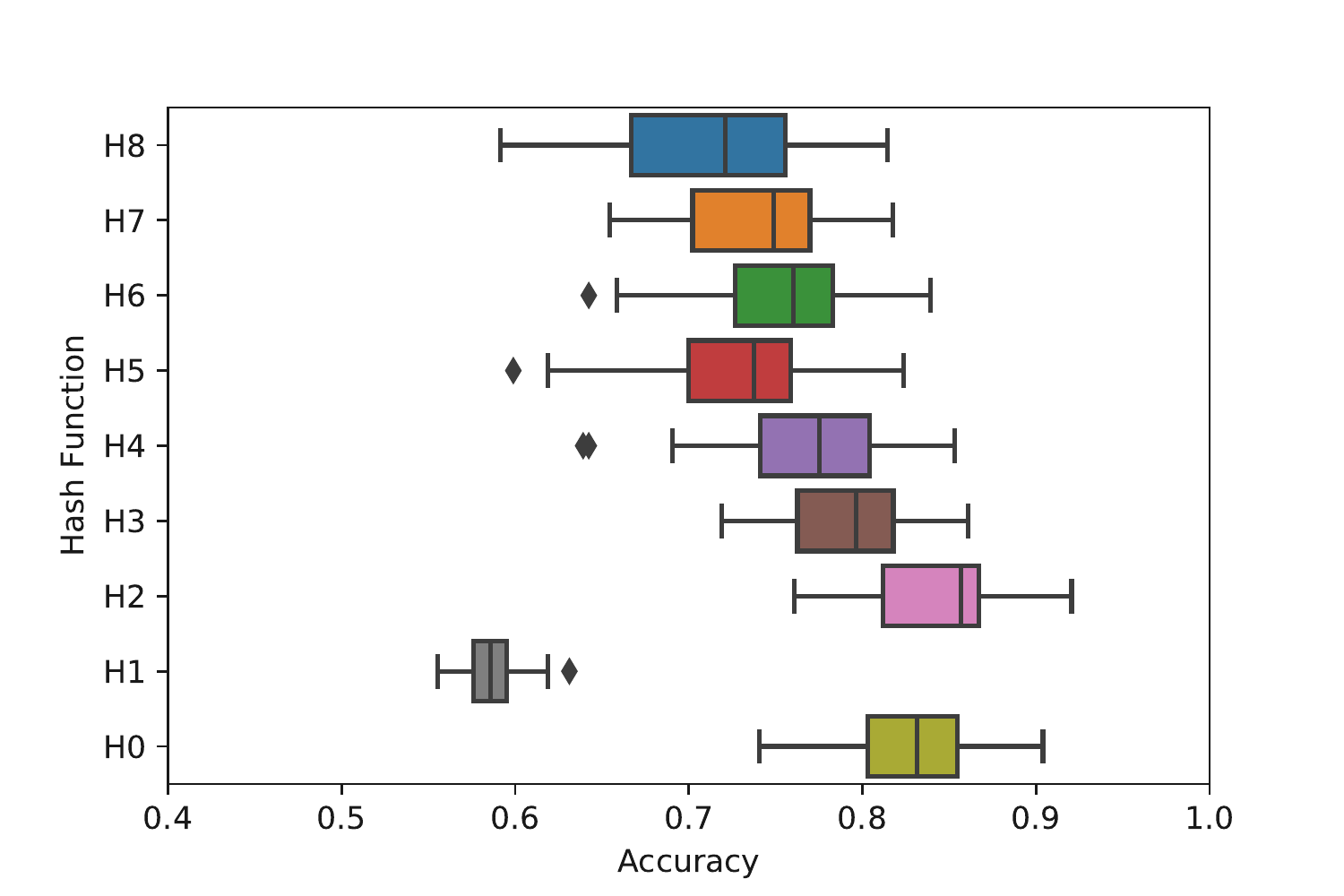}
\caption{Personalized DFAM classification accuracy for different mapping functions. $r=0.7, W=128,  g=3$.}
\label{maps}
\end{figure}

Figure \ref{maps} shows how the DFAM model performs across DFAM with different hash function implementations, in a personalized setting with windows of size $W=128$ samples, $g=3$ frequency bins, and a training window overlap ratio to be $r=0.7$. The vertical lines inside the colored blocks in Figure \ref{maps} represent the mean accuracy across individual datasets, and the lines to the left and right of each block corresponding to the first and third quartile of the distribution (individual dataset accuracies) with the blocks housing half of the distribution data points. The diamonds represent the outliers in accuracy results, which do not follow the distribution. We observe that hash functions $H0$ and $H2$ perform better than the remaining functions, and have mean classification accuracies of 0.78 and 0.8 with a standard deviation of 0.06 and 0.05, respectively. $H4$ and $H6$ closely follow them, with accuracies of 0.76 and 0.75, respectively. \emph{Based on these observations, we conduct further analysis on DFAM using the hash function $H2$ in the following sections}. 


\subsubsection{Bin Size and Window Size}
\label{sws}
Bin size ($g$) refers to the number of partitions dividing the sample frequency spectrum. The proposed DFAM uses \emph{g} frequency bins and extracts \emph{g} dominant frequencies (one from each bin to generate the dimension signature). To determine how the number of bins can affect the classification accuracy, we perform a combined analysis with different sized windows. Time-series data from smartphone and smartwatch sensors ($T_p$ and $T_w$) is divided into windows and each window contains $W$ motion sensor samples. Selecting the appropriate window sizes is critical for improving the overall responsiveness of the distraction detection framework. As the processing
time (and thus response time) is directly correlated to the window size, it is favorable to have a model that can detect activities using small windows sizes. A window of 32 samples can be obtained in approximately 0.64 seconds (at 50$Hz$), 64 samples in 1.28 seconds, and so on. However, we have to be careful in choosing $W$ because if the training window overlap ratio is high then the time taken to perform the activity matching increases as there will be more number of signatures to match against. 

\begin{figure}[]
\centering
\includegraphics[width=\linewidth]{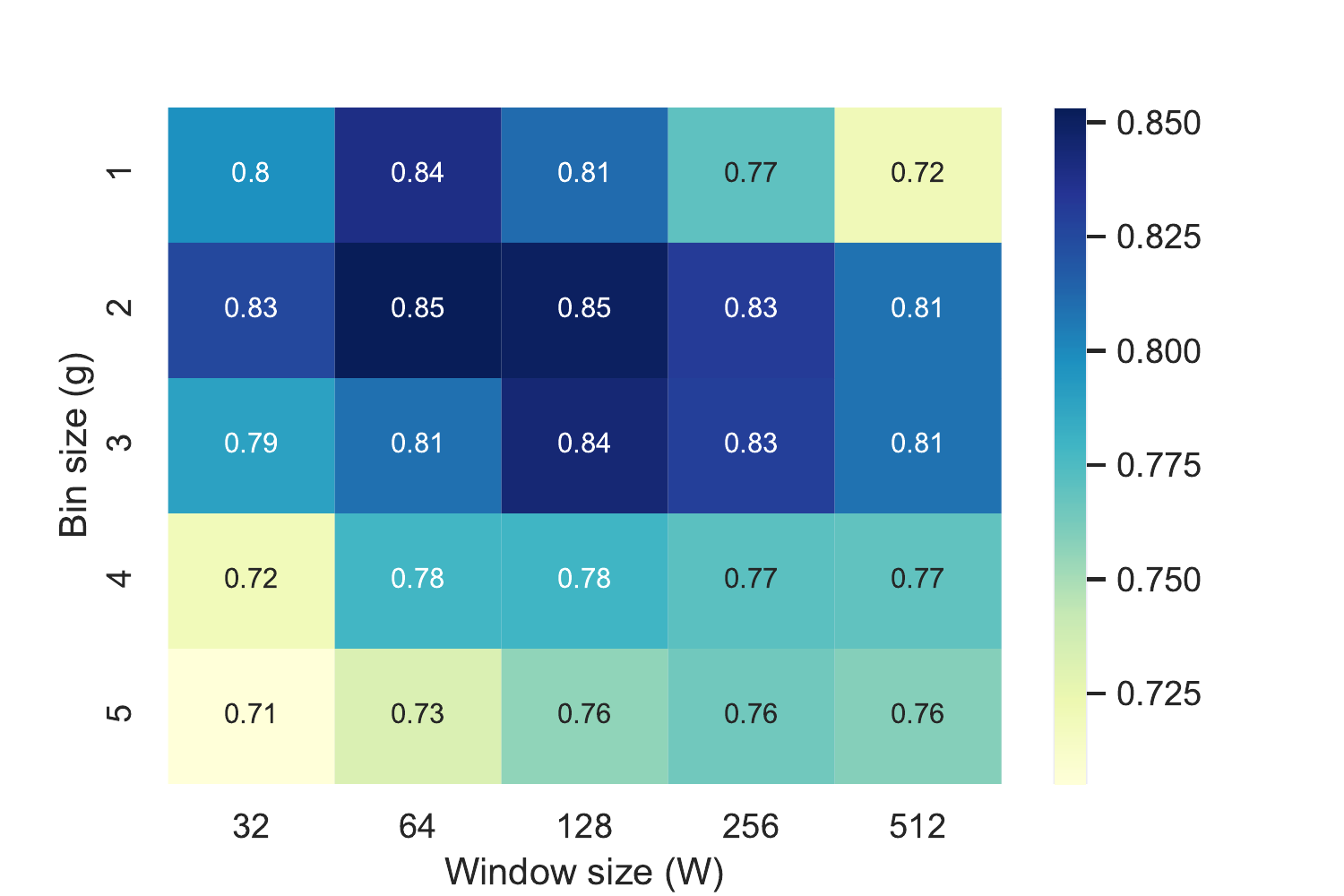}
\caption{Personalized DFAM classification accuracy for different window and bin sizes. $r=0.7$, $H=H2$.}
\label{winbin}
\end{figure}

Figure \ref{winbin} shows the performance of a DFAM model with a training window overlap ratio $r=0.7$ for different $g$ and $W$ combinations. We evaluated window sizes $W$ in powers of two in order to perform faster FFT computations. For a model with $g=3$ frequency bins, the mean accuracies of the trained classification model for 128 and 256 sample windows are 0.84 and 0.83, respectively. These values are higher than the mean accuracies of the models with a 32 sample window (0.79), indicating that the use of larger $W$ yields better results in most cases. The caveat to choosing a very large $W$ is that it takes more time to obtain a pedestrian activity, which may not be favorable for time-critical applications. \emph{Based on these constraints, we pick $W=128$ to be the window size of the DFAM model for the later evaluations}.

\subsubsection{Hardware}
Next, we analyze the performance of DFAM trained using the motion sensor data sampled from different make and model of smartwatches. The analysis will show whether DFAM trained on data from different smartwatches have comparable classification accuracies. There is a possibility that training and testing with data from one hardware yields better results than training and testing using data from another hardware. Given the availability of many smartphone and smartwatch models, it is important to know whether DFAM trained with data from different hardware have similar performance in a personalized setting. 

\begin{figure}[]
\centering
\includegraphics[width=\linewidth]{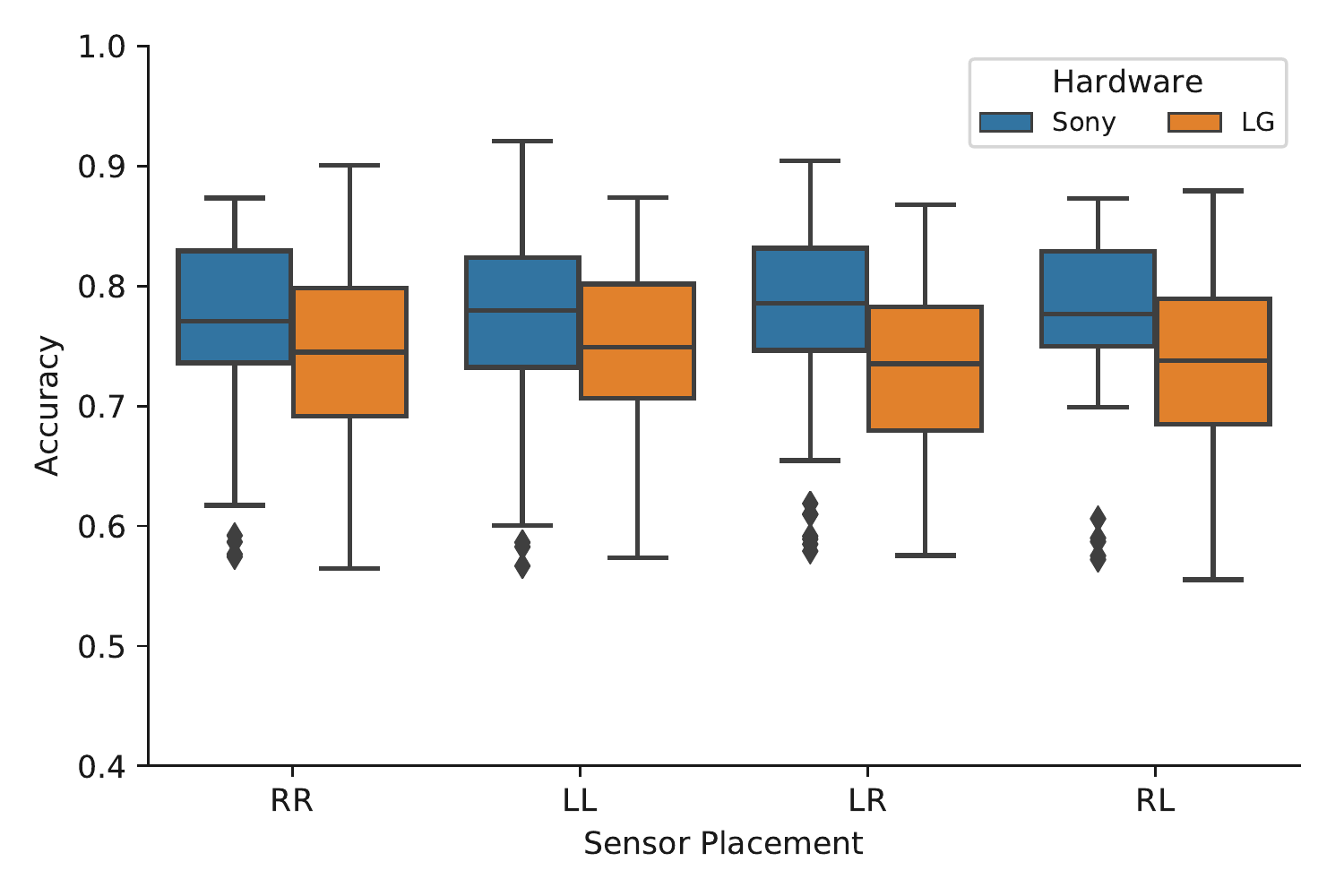}
\caption{Personalized DFAM classification accuracy for different hardware and sensor positions. $r=0.7$, $H=H2$, $W=128$, $g=3$.}
\label{device}
\end{figure}

Figure \ref{device} shows how much a DFAM model trained and tested with the data collected from LG smartwatch (Table \ref{datasetlist}) differs from a DFAM model trained and tested with data collected from Sony smartwatch. The comparison is done for each device placement scenario shown earler in Figure \ref{positions}. We can see that DFAM trained with Sony data has a mean accuracy of 0.78 accuracy, and performs slightly better than the DFAM trained with LG data, which has a mean accuracy of 0.74. Statistical analysis using a paired t-test \cite{finney1946statistical} shows that the DFAM models trained with Sony and LG data have comparable performance with t-statistic value of 2.00 and probability $p$ = 0.053. \emph{These observations indicate that DFAM models can have almost similar performance using hardwares from different manufacturers}.

\subsubsection{Device Placements}
In our experiments, same side device placements refer to scenarios where the smartwatch is worn on the same side as the smartphone (Figure \ref{positions}). For instance, we label the two possible same side scenarios as RR and LL, to denote left-left and right-right placement of the devices, respectively. Different side scenarios, on the other hand, represent scenarios where the smartwatch is worn on the opposite side of the smartphone. For example, in RL and LR placements, the smartwatch is worn on the right and left wrist, respectively, along with the smartphone in left and right pant pocket, respectively. Investigating these scenarios can provide light on the adaptability of the model, especially on whether the current DFAM model can still identify the pedestrian activities and perform consistently when sensor positions are changed. As for the preference for device placement differs from person to person, there may be cases where trained models from a certain device positioning may yield better results than others, which is why we consider the four common positionings shown in Figure \ref{positions}, inorder to find the best device placement among them.

\begin{figure}[]
\centering
\includegraphics[width=\linewidth]{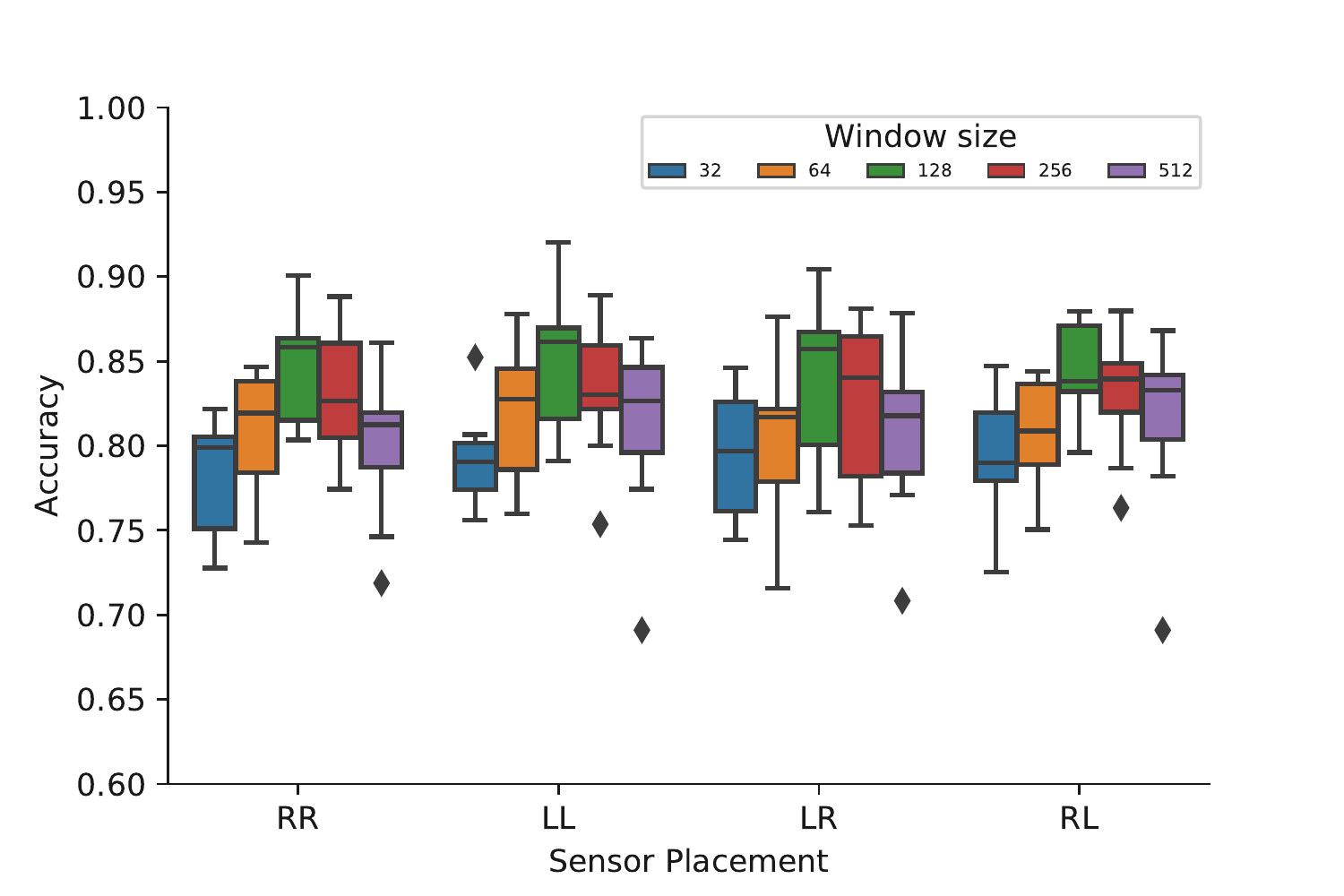}
\caption{Personalized DFAM classification accuracy for different sensor positions. $r=0.7$, $H=H2$, $g=3$.}
\label{place}
\end{figure}

In Figure \ref{place}, we consolidate the results across four different DFAM models for each of the four device-placement scenarios in the personalized setting. These placements are tagged based on the location of the smartwatch and smartphone. For a window with $W=128$ samples, the mean classification accuracy of a RR model (0.857) is very close to the mean classification accuracy of a LL model (0.861). We also observe the same side results average to 0.86 and is similar to the different side accuracies which averages to 0.85 for the same window size. \emph{Therefore, we can conclude that there is no difference in DFAM performance due to different device placements}. In addition to this, we perform a paired t-test \cite{finney1946statistical} to check if dominant hand placement yields better results than the non-dominant hand placement. The overall mean difference is close to zero with a t-statistic value of 0.39, and a $p$ value of 0.69. The small t-statistic value indicate that there is no significant performance difference between the DFAM model trained with activity data obtained from the dominant hand wrist and the model trained with non-dominant hand data, but the high $p$-value suggests a possibility that the results are obtained by chance. %

\subsubsection{Sensors}
The experiments in the previous sections used DFAM models trained using both accelerometer and gyroscope sensor data. In this section, we perform evaluations on two additional DFAM models: one model trained and tested using only accelerometer sensor data and the other model trained and tested using gyroscope data. This helps us determine whether a particular sensor or a combination time-series data is required (or beneficial) to precisely detect pedestrian related distracted activities. Such an analysis eliminates the collection and usage of extraneous or redundant data, and thereby reduce the response time of the distraction detection module by making classifications quick and efficient.  

\begin{figure}[]
\centering
\includegraphics[width=\linewidth]{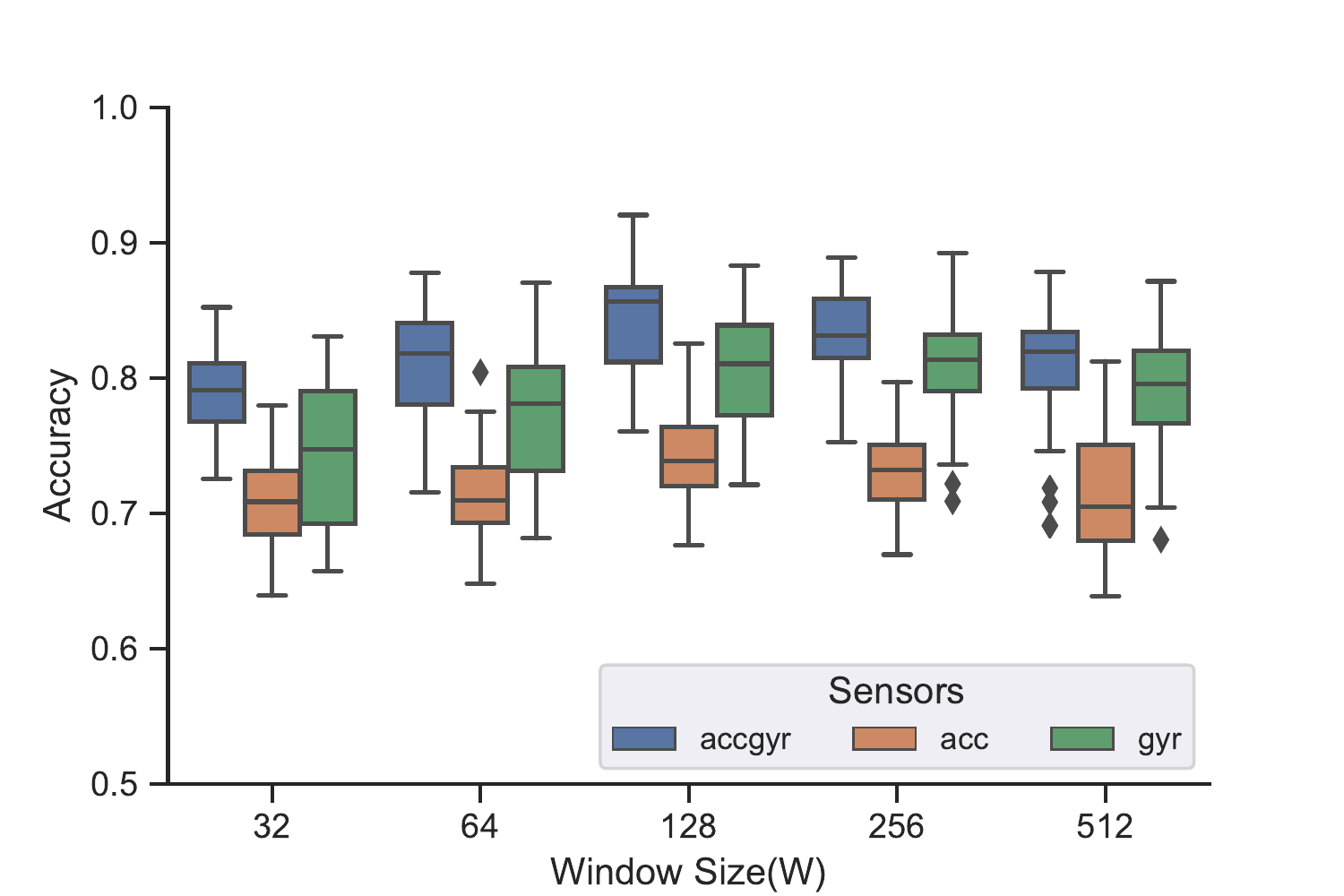}
\caption{Personalized DFAM classification accuracy for different sensors used in training. $r=0.7$, $H=H2$, $g=3$.}
\label{sensors}
\end{figure}

Figure \ref{sensors} shows the classification accuracies of DFAM models trained with different combinations of sensor data. Based on our empirical results, we observe an improvement in classification accuracy when DFAM model has access to both accelerometer and gyroscope data. In a single sensor setting, classification using the gyroscope data performs significantly better than using the accelerometer data. For instance, for a window of 128 samples, the mean accuracy of gyroscope-based classification models (0.81) is higher than the accelerometer-based models by 7.2\%, both of which are outperformed by the model with combined accelerometer and gyroscope data, having a mean accuracy of 0.86. \emph{Therefore, both accelerometer and gyroscope sensor data should be used by DFAM for achieving better distracted pedestrian detection accuracy.}

\section{Generalized Complex Activity Recognition}
The previous section evaluated the DFAM model performance in a personalized mode, where the participants provide their own data to train the models. However, some users may not be able to invest in the time and effort required for such a personalized training. Such cases require the use of a trained model (CAR) that generalizes well, that is able to classify the distraction related pedestrian activities from new users. In this section, we focus on designing and analyzing the performance of a generalized classification model.

\subsection{Generalizing DFAM}
\label{gcar}
A generalized CAR model is one which can correctly classify activities of new participants, without knowledge of their activity data. To design a generalized model, we created a multi-participant DFAM model by aggregating data from different participants and training using the combined data. We conduct experiments to analyze how effective this straightforward aggregation is in formulating a generalized CAR model. Incorporating data from multiple participants along with utilization of a high training overlap ratio $r$ and a sufficiently small window size ($W=128$) is likely to conceive a trained model encompassing a large number of activity signatures in the generalized than in the personalized setting. This is because the overlap ratio directly correponds to the number of training windows generated, and a large overlap ratio ($r$) may generate more windows for fixed-length time-series data when compared to a smaller $r$ value. As each window corresponds to a signature in the trained model, the number of signatures increases, and so does the number of matches done to recognize an unknown activity, which can increase the recognition time in the distraction detection module. Weighing down on these trade-offs, we trained models with a 20\% training window overlap ratio ($r=0.2$) for the generalized experiments instead of $r=0.7$ used in the personalized setting (Section \ref{dfamperformance}) to balance the overall framework performance. 

\begin{table*}[]
\centering
\caption{Generalized classification accuracy of DFAM and other classifiers. $r=0.2$, $H=H2$, $g=3$.} 
    \scriptsize
    \begin{tabular}{|r|rrrrrrrr|}
    \toprule
          & DFAM & SVM   & DT    & RF    & NB    & 1-NN   & 2-NN   & 3-NN \\
    \midrule
    W = 32 & 0.49 & \textbf{0.56} & 0.45 & 0.53 & 0.49 & 0.53 & 0.52 & 0.54 \\	
    W = 64 & 0.52 & \textbf{0.65} & 0.53 & 0.64 & 0.58 & 0.54 & 0.58  & 0.58 \\
    W = 128 & 0.61 & \textbf{0.67} & 0.55 & 0.65 & 0.58 & 0.58 & 0.58 & 0.59 \\
    W = 256 & 0.65 & 0.70 & 0.64 & \textbf{0.71} & 0.62 & 0.63 & 0.63 & 0.63 \\
    W = 512 & 0.69 & \textbf{0.76} & 0.64 & \textbf{0.76} & 0.67 & 0.69 & 0.69 & 0.70 \\
    \bottomrule
    \end{tabular}
  \label{dfamvsothers1}
\end{table*}

Table \ref{dfamvsothers1} shows the average classification measures of generalized CAR models, across placement scenarios from Figure \ref{positions}, using the Leave-One-Subject-Out Cross Validation (LOSO). In LOSO technique, one participant's dataset is used as the test set while the rest is used to generate the trained model, from a group of participants belonging to any placement scenario. The mean accuracy of these individual scenario models are computed for all CAR models, after which we compare the classification accuracies of DFAM with traditional CAR algorithms that are widely used for activity recognition in the literature, as shown in Table \ref{dfamvsothers1}. Results show that the accuracy of the DFAM model with window size $W=128$ is 0.61 and slighly lower that the mean classification accuracies of NB and SVM models which are 0.65 and 0.67 respectively. A repeated measures ANOVA test indicates that DFAM performance is significantly different from the traditional classifers based on the high F-statistic value of 13.50 and 7 degrees of freedom. The small probability value $p<0.001$ suggests that the results are less likely to be obtained by \textit{chance}. Although SVM has good classification accuracies for most windows sizes in Table \ref{dfamvsothers1}, DFAM is not far behind.

\subsection{On-Device CAR with Generalized DFAM}
\label{real-resource}
An on-device CAR evaluation is one in which the signature (or feature) generation and matching (or classification) is executed on the mobile and wearable device, unlike the evaluations in Sections \ref{dfamperformance} and \ref{gcar} which were executed on a PC. We implement the generalized models on the smartphone and smartwatch to analyze its response time and to measure the resources consumed by each CAR model. Quick response time is vital in determining the effectiveness of our framework, because any delay in alerting distracted pedestrians can be decisive in potential accident prevention. This should be accompanied by a lower resource utilization rate which decides the service time, or the amount of time the framework detects distracted activities and notifies the pedestrian.

\begin{table*}[]
\centering
\caption{Average resource consumption of Generalized Classifiers. $r=0.2$, $H=H2$, $W=128$.}
    \scriptsize
    \begin{tabular}{|r|rrrrr|}
    \toprule
	& Response & Utilization & Consumption & Utilization & Model  \\
	& Time (s)  & CPU (\%) & Power (mW)  & RAM (MB) & Size (KB)\\
    \midrule
     DFAM & \textbf{1.8} & 1.7\% & 33.3-129.5 & 37 & 236  \\
     SVM & 1.9 & 3.9\% & 33.3-188.7 & 43 & 229  \\
     DT & 1.9 & \textbf{0.8\%} & \textbf{33.3-85.1} & 36 & \textbf{111}  \\
     RF & 2.1 & 3.1\% & 85.1-222 & 68 & 6100  \\
     NB & \textbf{1.8} &  1.3\% & 40.7-96.2 & \textbf{20} & 131 \\ 
    1-NN & 1.9 &  2.1\%  & 85.1-214.6 & 23 & 1700  \\
    2-NN & 1.9 & 1.9\% & 85.1-188.7 & 32 & 1700 \\   
    3-NN & 1.9 & 2.1\% & 85.1-218.3 & 57 & 1700 \\
    \bottomrule
    \end{tabular}
\label{real-resource-table}
\end{table*}
Table \ref{real-resource-table} shows the DFAM implementation performance on the device, where we evaluate the response time, CPU, RAM and battery consumption of DFAM on the Motorola XT1096 smartphone paired with the Sony Smartwatch 3. The XT1096 with a 2300mAh Li-ion battery was running Android 6.0, while the Smartwatch 3 with a 420mAh Li-ion battery was running Android Wear 1.5. Results show that DFAM and NB have a lower response time (1.8$s$) for windows of 128 samples ($W=128$) compared to other traditional classifiers, which is beneficial for alerting distracted pedestrians in real-time. CPU utilization, power consumption and RAM utilization are also on the lower side, which means that users will notice a minimal impact on performance of their smartphone. With DFAM, we can update the existing trained model in real-time by simply appending the activity signatures computed to the model file. This manner of operation is not possible in case of NB models, where the models have to be re-trained in its entirety. \emph{These positive results position DFAM as the preferred candidate for use as a CAR technique in the distracted pedestrian detection framework}.

\section{Improving DFAM}
\label{idfam}
In this section, we further reduce the resource footprint and improve classification accuracy of the DFAM model from Section \ref{real-resource} by proposing a modification to the CAR framework through the introduction of a hierarchical model. We also implement the model on the device and measure its performance.

\subsection{A Hierarchical Model}
\label{har}

\begin{figure}[t]
\centering
         \includegraphics[width=0.85\linewidth]{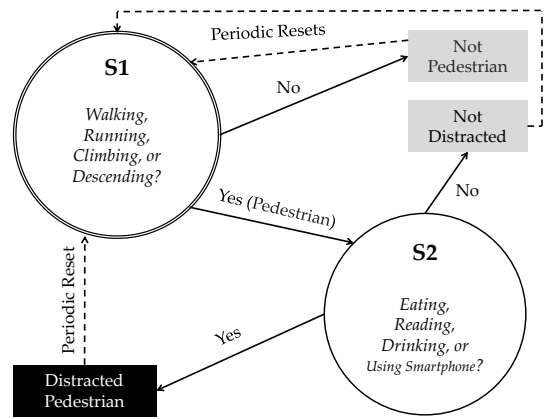}
	\caption{A hierarchical distracted pedestrian detection model.}
	\label{harfigure}
\end{figure}

Our hierarchical model is one in which high-level user activities are represented as states, as shown in Figure \ref{harfigure}. The framework design is based on the observation from Section \ref{sws} that smaller windows consume fewer resources to record and recognize an activity as opposed to larger windows for a fixed $r$. This coupled with the fact that most simple (repetitive) activities can be accurately recognized using windows with small $W$, and most complex activities require larger sample windows \cite{banos2014window,shoaib2016complex}, leads to a system model with two or more models: one with small $W$ for the simple activities and one (or more) with larger $W$ for the concurrent activities. The design is likely to achieve balanced performance outcomes, and thus have a positive impact on the overall user experience with the devices consuming lesser resources and providing extra service time. 

In the hierarchical model, when the system is in state S1, it is actively attempting to detect high-level moving activities from the device (smartphone+smartwatch) motion data. Once moving is detected, the system transitions to state S2 where it actively attempts to detect (and determine) the high-level distraction-related activity. If a distraction is not detected while the system is in state S2, the system transitions back to state S1. However, in case the pedestrian is detected to be distracted while in state S2, appropriate action is taken (e.g., user notification), and the system is set back to state S1 after an appropriate reset time. This reset time has to be carefully determined because if the reset is too frequent the user will obtain repetitive notifications, and users may in turn decide to not use our application anymore, or the notification itself can become a source of distraction. Further human factors study is required to determine the optimal duration between notifications to distracted users. Nonetheless, such a hierarchical CAR (HCAR) approach can reduce resource utilization and improve accuracy, especially when users are not mobile, because the activity recognition occurs within smaller sets of activities. 

It is also important to note that our hierarchical model simplifies the classification problem from a multi-class classification to a binary classification. For example, in state S1 the classification is to determine whether the user is a pedestrian or not, without regard for the type of pedestrian activity. Similarly, in state S2 the classification is to determine whether the user is distracted or not, without regard for the type of distracted activity. Intuitively, this simplified classification should improve the overall accuracy, compared to our previous multi-class classification results (Tables \ref{dfamvsothers1} and \ref{real-resource-table}).

\subsection{On-Device HCAR with Generalized DFAM} 
\label{hgcar}
The on-device HCAR is similar to the on-device CAR in Section \ref{real-resource} where the signature generation and matching are done on the smartphone and smartwatch with the exception of the hierarchical nature of the model.

We evaluate the accuracy, response time, CPU, and RAM utilization of the HCAR approach on the Motorola XT1096 smartphone paired with the Sony Smartwatch 3 using the pre-trained classification model from Section \ref{gcar} with parameters $r=0.2$, $H=H2$, $g=3$, and $W=128$. In this case the signature (or feature) generation and matching (or classification) is executed on the mobile and wearable device, unlike the evaluation in Section \ref{gcar} which was executed on a PC. The $All$ category in Table \ref{real-resource-table2} shows the DFAM implementation performance in a non-hierarchical model from Section \ref{real-resource}. We see significantly reduced resource utilization when users are in S1 (not pedestrian), that is, the CPU and RAM utilization significantly reduces from 1.7\% to 0.8\%, and from 37MB to 30MB, respectively in Table \ref{real-resource-table2}. Additionally, the combined classification accuracies of the generalized DFAM models at states S1 and S2 (0.79 and 0.74, respectively) in the hierarchical model in Table \ref{real-resource-table2}, is significantly better than the classification accuracy of the generalized model (0.61) in a non-hierarchical setting from Table \ref{real-resource-table}. Therefore, the hierarchical approach improves overall classification accuracy compared to our previous results in Tables \ref{dfamvsothers1} and \ref{real-resource-table}. \emph{These results validate that the hierarchal model can indeed be used within a pedestrian distraction detection framework, where it is critical to minimize the resource footprint of the framework without compromising its distracted activity detection accuracy and response time}.

\begin{table}[h]
\centering
\caption{Average resource consumption of HCAR with Generalized DFAM.}
    \scriptsize
    \begin{tabular}{|r|rrr|}
    \toprule
              & All & S1  & S2  \\
    \midrule
    Response Time & 1.8 s & 0.6 s & 0.9 s \\
    CPU Utilization & 1.7\%  & 0.8\%  & 1.5\%   \\  
    RAM Utilization & 37 MB  & 30 MB & 35 MB \\ 
    Power Consumption & 64.4 mW & 37.8 mW & 59.8 mW \\ 
    \bottomrule
    \end{tabular}
\label{real-resource-table2}
\end{table}

\section{Discussion}
\label{discussions}

Despite our really encouraging results, we feel that there were some limitations of our work, which we would like to address in our upcoming research efforts.

\subsection{Limitations} 
As our experiments were somewhat demanding for participants, they were unable to give longer (in terms of, time) datasets, which would have been preferable. The challenging nature of the experiments and the safety precautions also constrained the number/type of participants we could recruit. The smaller dataset size restricts us to analyze and compare results using more complex techniques, such as deep learning algorithms. However, our intuition is that while the use of deep learning algorithms may result in better classification accuracy, their high resource requirements may not be suitable for running on mobile and wearable devices.

\subsection{Future Work} 
As part of future work, we will be investigating into techniques that can further improve the current DFAM performance. We will also work towards developing an \textit{on-device alert module} for users' mobile and/or wearable devices that will remind distracted pedestrians that they should pay more attention to their surroundings while they are in motion. Additionally, we plan to implement a \textit{cloud-based alert module} that will employ crowd-sourced contextual information from distracted users to alert other users in the vicinity about the presence of distracted pedestrians. The design of these alert modules, however, is not trivial and requires a careful analysis of the associated human-factors issues. An alert mechanism that is not carefully designed may annoy users with frequent notifications, who may in turn decide to not use it anymore, or may itself become a source of distraction. We plan to accomplish this as part of our future work.
 
\section{Conclusion}
\label{conclusion}
We outlined and comprehensively evaluated a novel framework that detects and recognizes distracted pedestrian activities by using motion data available from users' mobile and wearable devices. As part of our framework, we designed and evaluated a novel dominant frequency matching based concurrent activity recognition model, called DFAM, and compared the performance and execution efficiency of the DFAM model with other well-known learning-based classification functions, such as Random Forests, SVM,  k-NN, Naive Bayes and Decision Trees. Our evaluation results showed that the proposed DFAM model is a suitable candidate for detecting concurrent activities, such as that of distracted pedestrians, and that it has reasonable concurrent activity recognition accuracy compared to traditional classification functions. We also observed that DFAM has lower power consumption rates and quicker response time(s) compared to Random Forests, SVM, k-NN, Naive Bayes and Decision Trees. In summary, we have not only comprehensively evaluated the efficacy and feasibility of various concurrent activity recognition techniques for detecting and recognizing pedestrian distraction, but have also proposed a novel concurrent activity recognition technique that achieves a good balance between recognition accuracy and alert response time, while being energy efficient. Finally, we proposed and evaluated a hierarchal model for recognizing distracted pedestrian activities, which further reduces resource utilization and improves detection accuracy.

\section{Acknowledgments}
Research reported in this publication was supported by the United States National Science Foundation (NSF) under award number 1829066 (previously 1637290).

\bibliographystyle{IEEETran}
\bibliography{apedsafe}

\clearpage
\newpage
\section*{Appendix A - Detailed Experimental Procedure}

{
\scriptsize


\noindent
\begin{tabularx}{\linewidth}{|X|}
    \toprule
    \textbf{Task Sets 1: Simple Activities} \\
    \midrule
    \textit{\textbf{Task A: Standing}} \\
    The participant will be standing in a sagittal posture; both arms at the sides. \\
    \textit{\textbf{Task B: Walking }} \\
    The participant will be walking along a hallway. The first minute in a slower pace, followed by normal and quick paces; each for a minimum duration of one minute. \\
    \textit{\textbf{Tasks C and D: Climbing Upstairs and Downstairs}} \\
    In these tasks, the participant will be climbing the stairs one at a time, without using the support of the side rails. The participant will initially be asked to climb down three floors of stairs inside a university building at a slow pace. After which, the participant will climb up the same stairs in a slow pace. Once the participant reaches the starting point (3rd floor), he/she will be urged to climb down and up again at the participant's usual pace. The step count and time taken to complete the tasks depends on individual participant's speed at which the activity is done. As combination of tasks C and D is more demanding, participants may take a short break (up to 3 minutes each) in between when they reach back at the starting point (3rd floor). \\
    \textit{\textbf{Task E: Sitting}} \\
    The participant will be sitting idle on a comfortable chair. \\
    \textit{\textbf{Task F: Running}} \\
    The participant will be running on the outdoor walkway. \\
    \bottomrule
\end{tabularx}


\noindent
\begin{tabularx}{\linewidth}{|X|}
    \toprule
    \textbf{Task Set 2: Concurrent Activities} \\
    \midrule
    \textit{\textbf{Task A: Walking + Using smartphone}} \\
    The participant will be asked to walk along a hallway, and use the smartphone to play a game (Tetris) at the same time. The smartphone is placed back in their pocket after this task is completed. \\
    \textit{\textbf{Task B: Walking + Reading}} \\
    The participant will be asked to walk while reading the front page of a printed newspaper (The Sunflower). \\
    \textit{\textbf{Task C: Walking + Eating}} \\
    The participant will be asked to walk and eat at the same time. Participant will be given a choice to pick the food they will be eating in this task. The choices will between a pack of chips, pretzels, or nuts. No restriction is imposed on the number of chips, pretzel, or nuts they eats during this task. After the task, the participant may eat any remaining food in the pack, or dispose it. \\
   \textit{\textbf{Task D: Walking + Drinking}} \\
    The participant will be drinking and walking simultaneously. Participant will be drinking from a bottle (17 fluid ounces) of drinking water at room temperature (about $70^\circ$F). After the task, the participant may drink any remaining water in the bottle, or dispose it. \\
    \textit{\textbf{Task E and F: Climbing Upstairs + Using Smartphone,  Climbing Downstairs + Using Smartphone}} \\
    In these tasks, the participant will be using the smartphone while climbing stairs (one at a time, without using the support of the side rails). The participant will be asked to climb down three floors of stairs   in Wallace Hall in their natural pace. The participant will then climb up the same stairs, while continuing their use of the smartphone. The step count and time taken to complete the tasks depends on individual participant's speed at which the activity is done. During these tasks, the participants will use the smartphone to play a game (Tetris). The smartphone is placed back in their pocket after this task is completed. \\
    \textit{\textbf{Task G and H: Climbing Upstairs + Reading, Climbing Downstairs + Reading}} \\
    In these tasks, the participant will be reading the front page of a printed newspaper (The Sunflower) while climbing stairs (one at a time, without using the support of the side rails). The participant will be asked to climb down three floors of stairs inside a university building at their natural pace. The participant will then climb up the same stairs, while continuing to read the newspaper. The step count and time taken to complete the tasks depends on individual participant's speed at which the activity is done. \\
    \textit{\textbf{Task I and J: Climbing Upstairs + Eating, Climbing Downstairs + Eating}} \\
    In these tasks, the participant will be eating (from a pack of chips, pretzels, or nuts) while climbing stairs (one at a time, without using the support of the side rails). The participant will be asked to climb down three floors of stairs inside a university building at their natural pace. The participant will then climb up the same stairs, while continuing to eat. The step count and time taken to complete the tasks depends on individual participant's speed at which the activity is done. After the tasks, the participant may eat any remaining food in the pack, or dispose it. \\
    \textit{\textbf{Task K and L: Climbing Upstairs + Drinking, Climbing Downstairs + Drinking}} \\
    In these tasks, the participant will be intermittently drinking water from a 17 fluid ounces water bottle, while climbing stairs (one at a time, without using the support of the side rails). The participant will be asked to climb down three floors of stairs inside a university building at their natural pace. The participant will then climb up the same stairs, while continuing to drinking water from the bottle. The step count time taken to complete the tasks depends on individual participant's speed at which the activity is done. After the tasks, the participant may drink any remaining water in the bottle, or dispose it. \\
\bottomrule
\end{tabularx}
\noindent
\begin{tabularx}{\linewidth}{|X|}
    \toprule
    \textbf{Concurrent Activities Continued ...} \\
    \midrule
    \textit{\textbf{Task M: Running + Using Smartphone}} \\
    The participant will be asked to run on outdoor walkway and use the smartphone at the same time. The participant will use the smartphone to play a game (Tetris). \\
    \textit{\textbf{Task N: Sitting + Using Smartphone}} \\
    The participant will sit on a comfortable chair and use the smartphone at the same time. The participant will use the smartphone to play a game (Tetris). \\
    \textit{\textbf{Task O: Standing + Using Smartphone}} \\
    The participant will stand and use the smartphone placed earlier in his/her pocket. The participant will then be asked to play a game (Tetris) for the entire time. The smartphone is placed back in their pocket after this task is completed. \\
    \textit{\textbf{Task P: Standing + Reading}} \\
    The participant will be given a print copy of the Wichita State University newspaper, The Sunflower. The participant will be asked to read an article on the front page in its entirety while standing. \\
    \textit{\textbf{Task Q: Standing + Eating}} \\
    The participant will be given a choice to pick the food they will be eating in this task. The choices will between a pack of chips, pretzels, or nuts. After picking one, the participant will be asked to stand and eat the food, one piece at a time. After the task, the participant may eat any remaining food in the pack, or dispose it. \\
    \textit{\textbf{Task R: Standing + Drinking}} \\
    The participant will be drinking a bottle (17 fluid ounces) of drinking water at room temperature (about $70^\circ$F) while standing. After the task, the participant may drink any remaining water in the bottle, or dispose it. \\
    \bottomrule
\end{tabularx}

}

\end{document}